\documentclass[preprint,authoryear]{elsarticle}

\usepackage[T1]{fontenc}
\usepackage{lmodern}
\usepackage{amsmath}
\usepackage{amssymb}
\usepackage{graphicx}
\usepackage{placeins}
\usepackage{algorithm}
\usepackage{algpseudocode}
\usepackage{multirow}
\usepackage[table]{xcolor}
\usepackage{hyperref}

\makeatletter
\def\input@path{{./}{}}
\makeatother

\graphicspath{{./}{figures/}}

\journal{Complexity}

\begin{document}

\begin{frontmatter}

\title{The Coordination Gap: Multi-Agent Alternation Metrics for Temporal Fairness in Repeated Games}

\author[uom]{Nikolaos Al. Papadopoulos\corref{cor1}}
\ead{nikolaos.papadopoulos@uom.edu.gr}
\fntext[orcid1]{ORCID: \url{https://orcid.org/0000-0003-1842-8227}}
\cortext[cor1]{Corresponding author}

\author[isir]{Ismael Tito Freire}
\ead{ismael.freire@sorbonne-universite.fr}

\author[upf]{Marti Sanchez-Fibla}
\ead{marti.sanchez@upf.edu}
\fntext[orcid3]{ORCID: \url{https://orcid.org/0000-0001-5725-1984}}

\author[uom]{Konstantinos E. Psannis}
\ead{kpsannis@uom.edu.gr}
\fntext[orcid2]{ORCID: \url{https://orcid.org/0000-0003-0020-6394}}

\address[uom]{Department of Applied Informatics, University of Macedonia, Egnatia 156, Thessaloniki 54636, Greece}
\address[isir]{Institut des Syst\`{e}mes Intelligents et de Robotique (ISIR), Sorbonne Universit\'{e}, Paris, France}
\address[upf]{Department of Information and Communication Technologies, Universitat Pompeu Fabra, Barcelona, Spain}

\begin{abstract}
Repeated multi-agent interactions require evaluation metrics that capture not only payoff distributions but also their temporal organization. Conventional outcome-based fairness measures can collapse temporally distinct coordination patterns into similar aggregate scores, obscuring whether access to a shared resource is genuinely rotating or persistently monopolized. We study this problem in the Honey-Jar Game (HJG), a minimally dynamic, repeated threshold-congestion Markov game in which $n$ agents compete for exclusive access to a single high-reward resource, serving here as a diagnostic benchmark for temporal coordination.

We introduce Perfect Alternation (PA), a reference coordination regime for turn-taking that corresponds to the $n$-periodic round-robin picking sequence, and propose six novel Alternation (ALT) metrics, together with a benchmarking methodology mapping ALT values to interpretable PA-equivalent performance. Using Q-learning agents as a minimal adaptive diagnostic baseline against random-policy baselines whose chance level we derive analytically, we uncover a clear measurement failure: despite deceptively high traditional metrics (e.g., reward fairness often exceeding $0.9$), learned policies fall below random baselines on every ALT variant, by $34\%$ to $74\%$ on CALT and by up to $92\%$ on EALT, with PA-equivalent coordination falling to roughly one-fifth of the population at $n=10$. Two qualitatively distinct failure patterns emerge depending on state representation, seen most sharply on EALT: without episodic memory (Type-A), the deficit grows monotonically from $-20\%$ at $n=2$ to $-92\%$ at $n=10$; with episodic memory (Type-B), a non-monotonic pattern emerges, with a trough at $n=5$ ($-76\%$) and partial recovery at $n=10$ ($-10\%$).

Traditional efficiency and fairness metrics do not reveal this distinction; the ALT framework serves as a diagnostic complement to the temporal fair division and picking sequence literature, addressing whether temporal fairness emerges spontaneously in decentralized adaptive systems rather than how to enforce it.
\end{abstract}

\begin{keyword}
Temporal Fairness \sep Congestion Games \sep Emergent Behavior \sep Agent-Based Modelling \sep Game Theory \sep Honey-Jar Game \sep Turn-Taking \sep Market-Entry Games \sep Repeated Games \sep Dynamic Games \sep Fair Division \sep Round-Robin Picking Sequence \sep Welfare Measurement \sep Social Choice
\end{keyword}

\end{frontmatter}


\section{Introduction}

Coordination among multiple self-interested agents is a fundamental problem in complex systems, where local decision rules and asymmetric incentives give rise to emergent collective behavior~\citep{rankin2007,axelrod1984,maruta2012,perolat2017}. Turn-taking versions of this problem arise across diverse domains: decentralized access in wireless networks~\citep{shenoy2007}, conversational turn-taking~\citep{sacks1974}, animal communication and rhythmic alternation~\citep{takahashi2013,greenfield2021,ravignani2019,pika2018}, stochastic phenotype switching~\citep{rainey2011}, and vision-based robotic coordination~\citep{mezey2025}. Yet existing evaluation frameworks, including welfare measures from social choice theory and allocation criteria from fair division, remain largely insensitive to the temporal structure inherent in these processes.

A growing body of work across computational social choice and decision theory has independently identified the same limitation: temporal fair division \citep{cookson2025,elkind2025,choi2026}, picking sequence frameworks \citep{bouveret2014,kalinowski2013}, perpetual voting \citep{lackner2020}, and temporally aware optimization \citep{torres2024} all demonstrate that cumulative fairness criteria are insufficient and that temporal structure must be explicitly accounted for across sequential allocation, voting, and planning settings. However, these works assume a central mechanism designer or planner who can enforce allocations; in contrast, our setting involves self-interested agents adapting through reinforcement learning with no central planner, raising the complementary question of whether temporal fairness can emerge spontaneously, and how to measure it when it does not.

Turn-taking tensions among many self-interested agents situate this problem within the broader family of congestion and anti-coordination games: Rosenthal's congestion games~\citep{rosenthal1973}, in which payoff depends on how many players share a resource; market-entry games~\citep{selten1982}, in which capacity-limited entry decisions collapse in value once demand exceeds supply; and the El Farol Bar Problem and Minority Game~\citep{arthur1994,challet1997}, which study the same anti-coordination tension for boundedly rational agents choosing repeatedly between two options. The two-agent Battle of the Exes (BoE)~\citep{hawkins2016} provides another minimal instance of this family, in which agents with asymmetric payoffs learn temporal alternation rather than static cooperation under repeated play~\citep{hawkins2016,papadopoulos2021,papadopoulos2020}; existing BoE studies, however, are almost exclusively restricted to the two-agent case~\citep{hawkins2016,puig2018,gasparrini2018,freire2020,freire2023,lopez2022}. Across this broader family, commonly used fairness and efficiency metrics remain time-averaged and payoff-based~\citep{gini1912,theil1967,allison1978,jiang2019,perolat2017}, rendering them insensitive to temporal ordering, and unable to distinguish genuine turn-taking from monopolistic, random, or weakly correlated access patterns.

In this study, we address these limitations by formalizing the Honey-Jar Game (HJG) as a Markov game~\citep{littman1994}. The game was introduced in our earlier work as the Multi-agent Battle of the Exes (MBoE), one possible, non-exclusive reading of its payoff structure inspired by BoE; as Section~\ref{sec:boe} makes precise, its minimally dynamic implementation with graded collision payoffs makes it, in essence, a multi-agent congestion game. We introduce Perfect Alternation (PA) as a reference coordination regime for idealized turn-taking and, building on this benchmark, propose six Alternation (ALT) metrics that explicitly capture temporal structure and coordination quality~\citep{papadopoulos2021,papadopoulos2025}. Through numerical simulations with Q-learning agents, used as a minimal and fully interpretable adaptive baseline rather than as a benchmark for optimal learning, we show that traditional metrics can mask coordination failure, while ALT metrics reliably distinguish qualitatively different long-run regimes.

\subsection{Main Contributions}

The main contributions of this study are as follows:

\begin{enumerate}
\item \textbf{Formalization of the Honey-Jar Game:} We formalize HJG, a BoE-inspired repeated congestion game for $n$ agents, as a Markov game, enabling the study of high-dimensional coordination dynamics.

\item \textbf{Perfect Alternation (PA):} We introduce PA as a reference coordination regime for turn-taking~\citep{papadopoulos2020,papadopoulos2021}, show that it corresponds to the $n$-periodic round-robin picking sequence~\citep{bouveret2014,kalinowski2013}, and that in HJG it achieves exact Temporal Envy-Freeness at every $n$-episode cycle boundary~\citep{choi2026}, being the maximally regular allocation with this property.

\item \textbf{Novel Alternation (ALT) metrics:} We propose six ALT measures that overcome the temporal insensitivity of traditional fairness and efficiency metrics, which cannot distinguish structured alternation from monopolistic or random access even when reporting values close to 1.0~\citep{papadopoulos2021,papadopoulos2025}. The framework serves as a diagnostic complement to the temporal fair division~\citep{cookson2025,elkind2025,choi2026} and picking sequence~\citep{bouveret2014,kalinowski2013} literature, assessing whether temporal fairness emerges spontaneously where no mechanism designer is present.

\item \textbf{ALT ratio benchmarking framework:} We develop a regression-based methodology mapping ALT metric values to Perfect Alternation--equivalent performance, enabling interpretable quantification of coordination quality.

\item \textbf{Random baseline as null process:} We establish random policy baselines, with an analytically derived null level, as a statistical reference for coordination, demonstrating that high fairness or efficiency values can arise purely by chance, a methodological practice not consistently reported in prior BoE and closely related coordination studies~\citep{hawkins2016,puig2018,gasparrini2018,freire2020,freire2023}.
\end{enumerate}

\subsection{Research Questions}

This study focuses on the following research questions:

\begin{enumerate}
\item[\textbf{RQ1:}] How do conventional fairness and efficiency metrics behave in multi-agent turn-taking settings, and under what conditions do they fail to distinguish structured coordination from random or monopolistic dynamics?

\item[\textbf{RQ2:}] How effectively do ALT metrics separate coordination regimes in empirical evaluation (perfect alternation, partial alternation, random access) in HJG?

\item[\textbf{RQ3:}] How does coordination quality, as measured by ALT metrics, scale with the number of agents in HJG, and does episodic state memory (Type-B) mitigate coordination failure at scale?

\item[\textbf{RQ4:}] How do ALT metrics compare against a random-policy null process, and what does this comparison reveal about apparent coordination measured by conventional metrics?
\end{enumerate}

The remainder of the manuscript is organized as follows: Section~\ref{sec:boe} introduces HJG; Section~\ref{sec:traditional} reviews traditional metrics and their limitations; Section~\ref{sec:alt} presents PA and the ALT metrics; Section~\ref{sec:experiments} describes the experimental methods; Section~\ref{sec:results} reports the results; Sections~\ref{sec:discussion} and~\ref{sec:conclusion} discuss the findings and conclude.

\section{The Honey-Jar Game and Related Coordination Games}\label{sec:boe}

\subsection{Congestion, Market-Entry, and Anti-Coordination Games}\label{sec:congestion_family}

A single resource whose value collapses under overuse situates this class of coordination problems within a broader family of congestion and anti-coordination games. Rosenthal~\citep{rosenthal1973} introduced congestion games, in which each player's payoff depends on how many others share the same resource, and showed that a potential function guarantees the existence of pure-strategy Nash equilibria. Market-entry games~\citep{selten1982} study a closely related setting in which $n$ players simultaneously decide whether to enter a market of limited capacity, with payoffs collapsing once the number of entrants exceeds it; experimental work in this tradition~\citep{rapoport1998} finds that aggregate entry converges near the capacity-efficient level even though individual behavior remains difficult to predict. The El Farol Bar Problem~\citep{arthur1994} and its formal abstraction, the Minority Game~\citep{challet1997}, capture the same anti-coordination tension for boundedly rational agents repeatedly choosing between two options.

\subsection{The Battle of the Exes Game}

The two-agent Battle of the Exes (BoE)~\citep{hawkins2016} provides another minimal instance of this family. It is a coordination game derived from the classical Battle of the Sexes, but with inverted incentives. Rather than coordinating on the same action, agents must avoid each other while competing for asymmetric rewards. If both agents select the same action, no reward is obtained; if they select different actions, each agent receives a payoff determined by individual preference asymmetry.

\begin{table}[h]
\centering
\caption{Two-player Battle of the Exes payoff matrix}
\label{tab:boe_payoff}
\begin{tabular}{l|cc}
\hline
 & \textbf{Action 1} & \textbf{Action 2} \\
\hline
\textbf{Action 1} & (0, 0) & (3, 2) \\
\textbf{Action 2} & (2, 3) & (0, 0) \\
\hline
\end{tabular}
\end{table}

BoE exists in two variants: a \emph{ballistic} formulation, in which agents make simultaneous discrete choices as in a normal-form game, and a \emph{dynamic} formulation, in which agents move toward preferred destinations in continuous space and time, so that approach trajectories are mutually observable before commitment. While the one-shot game admits multiple pure and mixed Nash equilibria, in repeated interaction the outcome combining full efficiency with equitable access is \emph{temporal alternation}: agents taking turns at the higher-reward outcome, a structured temporal pattern rather than a stationary payoff equilibrium. In experiments with human participants, Hawkins and Goldstone found that the dynamic formulation substantially increases the emergence of turn-taking, with some pairs converging to near-perfect alternation~\citep{hawkins2016}.

\subsection{Learning-Based Studies of BoE}

Following these findings, several studies examined whether artificial agents could reproduce the alternation behavior observed in humans. \citet{puig2018} investigated repeated BoE using independent Q-learning agents and reported convergence to stable deterministic strategies under specific learning-rate and exploration settings. \citet{gasparrini2018} incorporated loss aversion into agents' utility functions, accelerating convergence in the ballistic formulation and enabling alternation in the dynamic variant. Related modeling work explored control-based or episodic-memory agents in BoE-like settings, focusing on matching human efficiency/fairness/stability profiles rather than multi-agent scaling~\citep{freire2020,freire2023,lopez2022}. Despite these advances, learning-based BoE studies remain almost exclusively restricted to the two-agent setting, and random-policy baselines are not consistently reported as explicit null processes~\citep{hawkins2016,puig2018,gasparrini2018,freire2020,freire2023}, making it difficult to assess whether observed coordination reliably exceeds chance-level behavior. The exception is MSc thesis work within BoE itself, using randomized control comparisons in that same two-agent setting~\citep{lopez2022}, there to validate model performance against human data rather than as a formal null hypothesis. Our study addresses these gaps by (i)~formalizing random policies as an explicit statistical null process with a quantitative Coordination Score and Relative Change methodology, (ii)~scaling baselines to $n \in \{2,3,5,8,10\}$ across four experimental configurations, well beyond BoE's two-agent scope, and (iii)~demonstrating that traditional metrics can appear deceptively high even under random play. The paper builds on and substantially extends our early conceptual work on alternation-sensitive metrics~\citep{papadopoulos2020,papadopoulos2021,papadopoulos2025}.

\subsection{Why the Multi-Agent Setting Is Fundamentally Different}

With $n>2$ agents competing simultaneously for a single high-reward outcome, alternation is no longer binary role switching but collective periodic rotation across all participants; partial exclusion, intermittent monopolization, and irregular access patterns emerge naturally. Conventional fairness ratios also degrade structurally at scale, collapsing to extreme values that obscure intermediate agents (Section~\ref{sec:traditional}).

\subsection{Formalizing the Honey-Jar Game}

To address these limitations, we study an episodic Markov game involving $n \geq 2$ self-interested agents. We introduced this game in earlier work under the name \emph{Multi-agent Battle of the Exes (MBoE)}, a name we no longer use. The acronym is better read as \emph{Multi-agent Benefit of Exclusivity}, a more accurate description of its payoff structure and one possible, non-exclusive reading inspired by the two-agent game rather than a strict generalization of it. In each episode, agents act simultaneously and compete to reach a single high-reward terminal state. If exactly one agent succeeds, it receives the maximum payoff; if multiple but not all agents succeed, each receives a fixed reduced reward; if all agents tie, no reward is assigned. Payoffs are therefore non-increasing in the number of simultaneous claimants, a threshold-congestion structure with capacity collapse at full load. This implementation with graded collision payoffs makes the resulting game, in essence, a minimally dynamic, repeated threshold-congestion game; we therefore refer to it as the \emph{Honey-Jar Game} (HJG) throughout the remainder of this paper. The implementation is deliberately minimal (a single pre-terminal position) to maximize interpretability.

The name reflects a simple picture. A group of bear cubs, unable to communicate, share a single jar of honey. A cub dipping in alone gets a full paw of honey; if a few cubs dip in together, the opening narrows and each secures only a small fixed share; if all cubs rush the jar at once, the opening jams and nobody gets anything. Because each cub stands a few steps from the jar, approaching or holding back is visible to the others. Movement itself acts as a commitment signal, and the only collectively optimal behavior is spontaneous turn-taking.

The earlier two-agent game motivated this formulation but does not constrain it. Its payoff structure gives both agents a positive, merely asymmetric, reward whenever they choose different actions, whereas HJG is winner-take-all, with every non-winner receiving exactly zero in a given episode, and the dynamic formulation studied here unfolds as a multi-round race rather than a single simultaneous choice. Even at $n=2$, the two games are therefore not equivalent, and their equilibrium structures need not coincide; HJG stands on its own footing within the congestion and market-entry family introduced above, and Section~\ref{sec:congestion_diff} details how the two differ in payoff structure and evaluation focus. (HJG borrows its predecessor's ballistic/dynamic terminology for an analogous structural distinction, single decision per episode versus a multi-round race, without implying the underlying games coincide; see Section~\ref{sec:alt} and Supplementary Section~S4 for the ballistic case.) The optimal collective behavior in HJG is not simple avoidance but \emph{periodic turn-taking across all agents}. Whether and how such behavior emerges under decentralized learning (and how it should be evaluated) are the central questions addressed in the remainder of this manuscript.

We introduce Perfect Alternation as a reference coordination regime and propose novel, temporally sensitive Alternation (ALT) metrics capable of distinguishing genuine coordination from spurious access patterns.

\subsection{Differentiating HJG from Congestion, Market-Entry, and Anti-Coordination Games}\label{sec:congestion_diff}

Among the games introduced above, the market-entry game is the closest one-shot relative of HJG, since choosing \emph{move} is an entry decision toward a capacity-one prize with a zero outside option. Two structural features separate it from HJG nonetheless. First, the payoff structures respond to crowding differently. Entry payoffs typically decline with each additional entrant, whereas HJG's main experiments use a threshold step, a fixed per-capita share $r_{\mathrm{high}}/n$ for any partial tie, independent of how many agents tie, with full collapse only at unanimous entry (a per-claimant variant, $r_{\mathrm{high}}/k$ for $k$ simultaneous claimants, recovers the canonical single-resource congestion payoff exactly, and its quadratic counterpart $r_{\mathrm{high}}/k^{2}$ adds a commons-style degradation connecting HJG to common-pool resource settings~\citep{perolat2017}; both were explored in preliminary work). Second, the congestion and market-entry games introduced above are largely stateless and simultaneous-move, asking only how many agents enter, whereas HJG embeds the contest in a minimally dynamic environment in which agents' spatial approach toward the resource is observable before commitment, letting movement itself function as an implicit coordination signal, as Goldstone and Ashpole~\citep{goldstone2004} show for self-organizing access to a shared resource without explicit communication.

A more fundamental gap holds regardless of which reward rule governs collisions or how the contest is embedded in time. Even where this literature's own constructions repeat across rounds, as the El Farol Bar Problem and the Minority Game do, they track an aggregate statistic, whether the entry or choice rate converges near capacity, rather than a specific agent's identity over time; our evaluation asks the latter question directly, which agent obtains access and when across a repeated sequence.

An immediate question is whether these conclusions depend on the chosen reward rule. We address this directly rather than leave it open. The main experiments use the population-based scheme as an a priori design choice made for two reasons. It keeps the tie penalty deterministic and independent of the realized claim count, avoiding a confound between how many agents happen to collide in a given episode and the temporal-alternation signal ALT metrics target; and it keeps universal collision at exactly zero reward, so total non-coordination cannot register as partial success under outcome-based metrics such as efficiency and fairness measures, a property most visible at smaller $n$, where all agents colliding at once is common enough to noticeably shape the aggregate Efficiency score rather than being a rare edge case. A robustness check with the per-claimant split (Supplementary Section~S5) shows the coordination gap persists at every population size tested and widens only at $n=2$, for the structural reason that every collision there involves all agents; the finding is therefore not an artifact of the reward-threshold choice. The same check also makes a second, independent point. Because this per-claimant scheme is literal Rosenthal congestion, standard congestion-game equilibrium analysis does not detect the failure, while the ALT/PA framework does. This is the evaluation-focus gap from above resurfacing inside a canonical congestion payoff.

\subsection{The Honey-Jar Game as a Markov Game}
\label{sec:mboe_markov}

We now formally define the \emph{Honey-Jar Game} (HJG) as an episodic Markov game~\citep{littman1994}. This formulation is a core contribution of this work: it embeds a winner-take-all contest for a single resource in a scalable congestion setting, motivated by but structurally distinct from the two-agent BoE.

\subsubsection{Game Definition}

HJG is defined as a Markov game
\begin{equation}
\mathcal{G} = (N, S, A, T, R),
\end{equation}
where $N = \{1,2,\dots,n\}$, $n \geq 2$, is the set of self-interested agents, each maximizing its own cumulative reward, $S$ is the state space, $A$ the joint action space, $T$ the transition function, and $R$ the reward function. The game is episodic. Each episode begins from a fixed initial configuration and terminates when at least one agent reaches a terminal state; in the implementation, episodes are additionally truncated after 100 rounds if no agent has reached the terminal, and truncated episodes count as no-winner episodes for all metrics. At each time step $t$, every agent $i \in N$ simultaneously and independently selects an action
\begin{equation}
a_t^i \in \mathcal{A} = \{\text{move}, \text{stay}\},
\end{equation}
with joint action $\mathbf{a}_t = (a_t^1,\dots,a_t^n)$. In the minimally dynamic environment these actions correspond to temporal strategies: \emph{move} commits the agent one step closer to the contested resource, while \emph{stay} yields, holding position to avoid collision. The transition function $T : S \times A \rightarrow S$ is deterministic and updates each agent's position according to
\begin{equation}
p_{t+1}^{i} =
\begin{cases}
p_t^i + 1, & \text{if } a_t^i = \text{move}, \\
p_t^i, & \text{if } a_t^i = \text{stay},
\end{cases}
\end{equation}
so positions are non-decreasing and advance by at most one per step.

\subsubsection{Reward Function}

Rewards are assigned only at terminal states. Let
\begin{equation}
k = \sum_{i=1}^n \mathbb{I}(p_T^i = \text{terminal})
\end{equation}
denote the number of agents reaching the terminal position at the end of an episode. The reward for agent $i$ is given by
\begin{equation}
R_i(s_T,\mathbf{a}_T) =
\begin{cases}
r_{\mathrm{high}}, & \text{if } k = 1 \text{ and agent } i \text{ succeeds alone}, \\
r_{\mathrm{low}}, & \text{if } 1 < k < n \text{ and agent } i \text{ succeeds}, \\
0, & \text{otherwise}.
\end{cases}
\end{equation}

If all agents reach the terminal position simultaneously ($k=n$), no reward is assigned. This abstracts the two-agent BoE's asymmetric destination preferences into a single contested terminal with exclusivity-based rewards, preserving the core tension (benefit from exclusive access, interference under competition) while scaling to arbitrary $n$. For instance, in a 3-agent system with $r_{\mathrm{high}} = 100$: if agent 1 wins alone, $R_1 = 100$ and $R_2 = R_3 = 0$; if agents 1 and 2 both reach the terminal (partial tie), $R_1 = R_2 = r_{\mathrm{low}}$ and $R_3 = 0$; if all three agents tie, $R_1 = R_2 = R_3 = 0$. Collectively efficient and temporally equitable behavior corresponds to \emph{periodic turn-taking} across episodes, the basis for the Perfect Alternation reference regime of Section~\ref{sec:alt}.

\section{Traditional Multi-Agent Fairness Measures}\label{sec:traditional}

Multi-agent fairness research distinguishes between \emph{outcome fairness} (cumulative payoff distribution) and \emph{procedural fairness}~\citep{suzumura2011} (temporal distribution of rewards). Common outcome measures include the Gini coefficient~\citep{gini1912}, Theil index~\citep{theil1967}, and Coefficient of Variation~\citep{allison1978}, while recent multi-agent work often uses payoff-based fairness or equality variants~\citep{jiang2019,perolat2017}. In the fair division literature, allocation mechanisms are evaluated by properties such as envy-freeness, proportionality, and strategyproofness~\citep{suksompong2023,camacho2023,darmann2016}, yet these criteria remain distributional and do not account for the temporal sequence of allocations. Even sequential allocation models that explicitly incorporate agent ordering, as in river sharing problems~\citep{ansink2012}, focus on the distributional fairness of final shares rather than on the dynamic temporal pattern of access. Similarly, axiomatic approaches to cooperative surplus sharing emphasize distributional fairness and consistency~\citep{calleja2020}, but do not address the temporal ordering of payoffs. This limitation has been independently recognized in the temporal fair division literature: \citet{cookson2025} show that per-round cumulative criteria are insufficient without explicit temporal guarantees, and \citet{elkind2025} prove that Temporal Envy-Freeness up to one item (TEF1) is incompatible with Pareto optimality in general, confirming that standard efficiency and fairness criteria jointly fail to characterize temporal coordination structure even from a mechanism design perspective. None of these criteria captures \emph{when} agents receive rewards; see Supplementary Section~S3 for extended discussion. Axiomatic treatments of temporal fairness exist in adjacent domains, such as scalar fairness measures for realized service order in queueing systems~\citep{avi-itzhak2008} and characterizations of which tournament formats guarantee equal winning chances in round-robin-based league competitions~\citep{arlegi2023}, but neither targets a scalar measure of alternation quality for a repeated single-resource contest, the gap the ALT family addresses.

\subsection{Limitations of Traditional Metrics in BoE and HJG}

We examine four traditional metrics and their shortcomings for the Battle of the Exes (BoE) and the Honey-Jar Game (HJG). In what follows, $w_i$ counts exclusive wins, $t_i$ counts terminal arrivals (including ties), and $p_i$ is the cumulative payoff of agent $i$ over $\nu$ episodes.

\paragraph{Fairness.}
Defined as $F^{\nu} = \min_i w_i / \max_i w_i$, taken over the agents recording at least one exclusive win. Key limitations: (1) ignores intermediate agents (only min/max), (2) a completely excluded agent leaves the ratio entirely, so shutting out one or more agents does not by itself lower the score, and (3) identical values arise from fair rotation, subgroup monopoly, or random access.

\paragraph{Efficiency.}
Defined as $E^{\nu} = \sum_i r_i^{\text{total}} / (\nu \times r_{\text{high}})$, measuring total reward capture. Key limitations: (1) high efficiency can result from monopoly or alternation, and (2) provides no information about coordination patterns.

\paragraph{Turn-Taking Fairness.}
Defined as $F_{\nu}^{TT} = \min_i t_i / \max_i t_i$~\citep{hawkins2016}, again over agents with positive counts. Key limitations: (1) still ignores intermediate agents, and (2) high $t_i$ can reflect frequent collisions rather than genuine alternation.

\paragraph{Reward Fairness.}
Defined as $F_{\nu}^{R} = \min_i p_i / \max_i p_i$~\citep{hawkins2016,puig2018,gasparrini2018}, over agents with positive payoff, comparing cumulative payoffs. Key limitations: (1) high values can result from tie rewards without turn-taking, and (2) a dominant subgroup can still yield high Reward Fairness.

All three ratios are computed over agents with positive counts, since applied literally across all $n$ agents each collapses to zero the moment a single agent is shut out, however evenly the remaining $n-1$ share access. Two degeneracies follow, both worked through in Supplementary Section~S3. The restriction itself hides total exclusion, and at the opposite extreme the denominator vanishes, because a unanimous tie pays every agent zero under the HJG reward rule, so a run of nothing but full collisions leaves $\max_i p_i = \max_i w_i = 0$ and the ratios undefined at the exact point of total coordination collapse. Reading equal shares as perfect equity would return $1$ there, every agent having received an identical amount of nothing, and $F_{\nu}^{TT}$ reaches that verdict with no convention at all, reporting perfect turn-taking fairness for a system with zero reward and zero coordination. Awarding a maximal score to complete collapse is not temporal blindness but inversion.

\begin{table}[t]
\centering
\caption{Diagnostic limitations of traditional metric combinations. Each pairing was systematically evaluated across all nine value combinations arising from Efficiency $\in \{0, (0,1), 1\}$ and the respective Fairness variant $\in \{0, (0,1), 1\}$. Complete cell-by-cell analysis appears in Supplementary Section S3 (Tables 2 to 4).}
\label{tab:combined_metrics}
\footnotesize
\resizebox{\columnwidth}{!}{%
\begin{tabular}{l|c|c|p{4.5cm}}
\hline
\textbf{Metric Pair} & \textbf{Temporal} & \textbf{Fair Outcome} & \textbf{Key Failure Mode} \\
\textbf{(with Efficiency)} & \textbf{Structure} & \textbf{Confirmed?} & \\
\hline
Fairness & None & No & Considers only min/max agents; ignores intermediate distribution and temporal sequence \\
Turn-Taking Fairness & None & No & Collision inflation: ties increment all agents' counts equally, conflating arrival with turn-taking \\
Reward Fairness & None & Only at (E=1, RF=1) & Payoff-based; high values achievable via ties alone; temporal sequence entirely invisible \\
\hline
\end{tabular}%
}
\end{table}

Combinations do not repair the problem: Fairness ignores intermediate agents, TT Fairness conflates exclusive wins with collision-induced arrivals, and even Reward Fairness combined with high Efficiency leaves temporal structure invisible. Table~\ref{tab:combined_metrics} shows that interpretable patterns emerge only at extremes; at the intermediate values common in practice, these metrics cannot distinguish coordination quality.

\section{Proposed Assessment Framework}\label{sec:alt}

Outcome-based metrics aggregate payoffs across episodes, losing information about \emph{temporal structure} (who succeeds, when, and in what sequence). ALT metrics explicitly quantify these patterns.

\subsection{Perfect Alternation as Reference}

\textbf{Perfect Alternation (PA)} is a coordination regime where each of $n$ agents reaches the highest-payoff state exactly once within every sliding window of $n$ consecutive episodes, in any order (equivalently, the outcome sequence is $n$-periodic). Stricter order-sensitive variants (e.g., requiring each agent to occupy each rank equally often across batches) are possible but not imposed here. PA can be used descriptively, normatively, or prescriptively depending on the application; here it serves only as a reference regime against which observed coordination can be compared~\citep{papadopoulos2020,papadopoulos2021}. Three reasons motivate this choice. First, \emph{empirical observability}: human participants in BoE experiments sometimes converge to PA-like patterns over extended interaction~\citep{hawkins2016}, so PA is an attainable regime rather than a purely theoretical construct. Second, \emph{temporal distribution sensitivity}: standard fairness measures are insensitive to when rewards are obtained, whereas PA is the extreme case in which temporal disparities are minimized by construction. Third, \emph{metric calibration}: fixing a coordination pattern of maximal regularity provides a natural normalization point for alternation-sensitive metrics, anchoring $\mathrm{ALT}=1$ to PA. PA is Pareto-optimal under strict rotation, and in the two-agent ballistic case of HJG (a single simultaneous move/stay decision per episode, rather than the multi-round race studied experimentally) it constitutes a Nash equilibrium; detailed equilibrium links are provided in Supplementary Section~S4.

The HJG setting corresponds precisely to the identical days case with house allocation in temporal fair division \citep{cookson2025,elkind2025}: one indivisible good (the high-reward state) arrives each episode, and exactly one agent receives it. In this setting, Perfect Alternation corresponds to the $n$-periodic round-robin picking sequence \citep{bouveret2014,kalinowski2013}, which maximizes expected utilitarian welfare for two agents and, by Theorem~16 of \citet{choi2026}, guarantees that every $n$ episodes each agent has won exactly once, achieving not merely TEF1 but exact Temporal Envy-Freeness (TEF) at each cycle boundary. Adams and Segal-Halevi~\citep{adams2026} study the closely related problem of repeatedly assigning $n$ items to $n$ agents through balanced sequences of permutations, in which each agent receives each item exactly once per $n$-round cycle (a Latin-square structure); PA is precisely this structure specialized to a single contested good. Under the symmetric single-resource valuation of HJG, PA is thus sufficient for temporal proportionality and for exact envy-freeness at $n$-episode cycle boundaries, but it is not the unique such outcome: any allocation that gives each agent one win per disjoint block is proportional and cycle-boundary envy-free without being perfectly alternating (for $n=2$, the clumped sequence $A,B,B,A,A,B,B,A,\ldots$ is one example). PA is distinguished by an additional regularity requirement, constant inter-win gaps equal to $n-1$, and is therefore the maximally regular member of the temporally proportional class rather than the unique envy-free allocation. This provides independent theoretical grounding for PA as a natural welfare benchmark in our setting.

\subsection{ALT Framework}

Given $\nu$ episodes and $n$ agents, ALT metrics evaluate all $b = \nu - (n-1)$ overlapping batches of $n$ consecutive episodes. Each batch $j$ receives a weight $\beta_j \in [0,1]$, where $\beta_j = 1$ indicates perfect alternation. The overall score averages batch weights:
\begin{equation}
\mathrm{ALT} = \frac{1}{b} \sum_{j=1}^{b} \beta_j
\end{equation}

ALT measures satisfy: (i) boundedness in $[0,1]$; (ii) $\mathrm{ALT}=1$ under PA; (iii) monotonic degradation under monopolization or ties; (iv) sensitivity to exclusivity; (v) permutation invariance~\citep{papadopoulos2021,papadopoulos2025}. For this study, order within a batch is not penalized; any permutation is treated as ideal. Full symbol definitions and computation details are provided in Supplementary Section~S4.

Three independent checks calibrate what the metrics measure, none of which involves learned behavior: a positive control (synthetic PA sequences, for which every variant returns $1$), graded-response sweeps (the AltRatio benchmarking simulations for $n \in [2,40]$, Supplementary Section~S1, and a clumping sweep at constant win shares, Supplementary Section~S4, in which metric values respond monotonically as alternating structure degrades), and a theoretical null (the analytic random model of Section~\ref{sec:random_baselines}, whose predicted chance levels match the measured baselines). Metric validity therefore rests on constructed sequences with known properties and on probability theory, not on the Q-learning results that the metrics are subsequently used to evaluate.

\subsection{Six ALT Variants}

We define six variants organized into three primary and three secondary metrics. For batch $j$, let $f_j$ denote the number of distinct agents reaching the terminal state at least once; $\tau_j$ the total number of terminal arrivals; $w_j$ the number of episodes with an exclusive winner; $g_j$ the number of agents with exactly one exclusive win; and $Y_{j,\ell}$ the number of agents reaching the terminal in the $\ell$-th episode of the batch (written $Y_\ell$ when the batch is clear from context).

\subsubsection*{Primary Metrics}

\paragraph{CALT (Complete)} Primary metric; jointly accounts for tie frequency and winner diversity by weighting the fractional score with an explicit tie penalty:
\begin{equation}
\beta_j^{\mathrm{CALT}} = \frac{\sum_{\ell=1}^{n} (n - Y_\ell) \cdot \beta_j^{\mathrm{qFALT}}}{n(n-1)}
\end{equation}
where $\beta_j^{\mathrm{qFALT}} = (f_j/\tau_j)^2$ is the squared fractional score (introduced with the secondary metrics below) and $(n - Y_\ell)$ is the per-episode tie penalty (number of agents \emph{not} reaching a terminal state in episode $\ell$). Full derivations and sensitivity analysis are provided in Supplementary Section~S4.

\paragraph{EALT (Exclusive)} Rewards batches in which agents win exclusively (single-agent episodes), weighting both the number of such episodes and the diversity of winners:
\begin{equation}
\beta_j^{\mathrm{EALT}} = \frac{w_j \cdot f_j}{n^2}
\end{equation}

\paragraph{AALT (Absolute)} Strictest primary metric; awards credit only to agents that achieve exactly one exclusive win per batch, requiring near-perfect rotation:\footnote{AALT values throughout this version were recomputed with a corrected implementation. An earlier version also credited tie arrivals toward $g_j$. All qualitative conclusions are unchanged, and every coordination score remains negative.}
\begin{equation}
\beta_j^{\mathrm{AALT}} = \frac{g_j}{\tau_j}
\end{equation}

\subsubsection*{Secondary Metrics}

\paragraph{FALT (Fractional)} Most tolerant metric; the proportion of unique winners relative to total terminal occurrences, $\beta_j^{\mathrm{FALT}} = f_j/\tau_j$, without penalizing ties or overlap, making it a loose lower bound that is insensitive to monopolization. The quadratic variants qFALT and qEALT square the corresponding batch weights, $\beta_j^{\mathrm{qFALT}} = (\beta_j^{\mathrm{FALT}})^2$ and $\beta_j^{\mathrm{qEALT}} = (\beta_j^{\mathrm{EALT}})^2$, amplifying sensitivity near perfect coordination.\footnote{Previously called EFALT/EEALT; renamed since the operation is quadratic, not exponential.}

\noindent CALT serves as the primary metric due to its balanced treatment of exclusivity and ties. EALT isolates exclusive access, and AALT sets a strict diagnostic bound. Secondary metrics provide supplementary perspectives reported in the Supplementary Material. ALT measures complement (rather than replace) traditional metrics by capturing temporal coordination invisible to outcome-based evaluation; we recommend reporting them jointly rather than aggregating them into a single index, since Efficiency and Reward Fairness describe how much is gained and how evenly, while ALT describes \emph{when}. The three-stage interpretive pipeline (CALT, then PA-equivalent, then Coordination Score) is illustrated in Supplementary Section~S3.

Distance-based alternatives, such as a Kendall-tau distance between the observed win sequence and a reference ordering, are possible in principle but fit this setting poorly: PA is not a single reference sequence but an equivalence class (any within-window permutation at any phase), ties and repeated winners fall outside permutation distances, and sliding-window evaluation localizes deviations in time instead of aggregating them into one global distance. The AltRatio mapping of Section~\ref{sec:experiments} then supplies the intuitive ``percentage of perfect alternation'' reading on top of the window-level measurements rather than in place of them.

\paragraph{Worked example} Consider $n=3$ agents $\{A,B,C\}$ and $\nu=6$ episodes with outcome sequence $A,\, B,\, C,\, \{A{,}B\},\, C,\, A$, where $\{A{,}B\}$ denotes a tie episode in which both $A$ and $B$ reach the terminal state. The $b = 6-2 = 4$ sliding windows of three consecutive episodes receive the following weights (computed with the reference implementation):
\begin{table}[h]
\centering
\caption{Worked example: batch weights for the six-episode sequence $A,B,C,\{A,B\},C,A$ across the four sliding windows.}
\label{tab:worked_example}
\small
\resizebox{\columnwidth}{!}{%
\begin{tabular}{lcccc|c}
\hline
Window & $(A,B,C)$ & $(B,C,\{A{,}B\})$ & $(C,\{A{,}B\},C)$ & $(\{A{,}B\},C,A)$ & Mean \\
\hline
$\beta^{\mathrm{FALT}}$ & 1.000 & 0.750 & 0.750 & 0.750 & 0.812 \\
$\beta^{\mathrm{CALT}}$ & 1.000 & 0.469 & 0.469 & 0.469 & 0.602 \\
$\beta^{\mathrm{AALT}}$ & 1.000 & 0.500 & 0.000 & 0.500 & 0.500 \\
\hline
\end{tabular}%
}
\end{table}
The first window is perfectly alternating, so every metric assigns weight $1$. In the second window the tie inflates $\tau_j$ to $4$ while all three agents still reach the terminal state ($f_j = 3$): FALT degrades only mildly ($3/4$), CALT additionally applies the per-episode tie penalty ($\beta = 5 \times 0.5625 / 6 = 0.469$), and AALT credits only the two agents with exactly one exclusive win ($g_j/\tau_j = 2/4$). The third window isolates AALT's strictness. Agent $C$ wins exclusively twice, so no agent has exactly one exclusive win and $\beta^{\mathrm{AALT}} = 0$, while FALT is unchanged at $0.750$. A step-by-step walkthrough of all six variants on this sequence is provided in Supplementary Section~S4.

\section{Methods}\label{sec:experiments}

\subsection{Environment and Experimental Design}

We study HJG coordination using two state representations. \textbf{Type-A} (position-only: $s_t^{A} = \mathbf{p}_t$, with $\mathbf{p}_t = (p_t^1,\dots,p_t^n)$) and \textbf{Type-B} (with episodic memory: $s_t^{B} = (\mathbf{p}_t, \mathbf{z}_{t-1})$, with $\mathbf{z}_{t-1} = (z_{t-1}^1,\dots,z_{t-1}^n)$ and $z_{t-1}^i=1$ if agent $i$ won the previous episode). Two reward schemes scale tie penalties. \textbf{ILF} (Inverse Linear Fractional; $r_{\mathrm{low}} = r_{\mathrm{high}}/n$) and \textbf{IQF} (Inverse Quadratic Fractional; $r_{\mathrm{low}} = r_{\mathrm{high}}/n^2$), with $r_{\mathrm{high}}=100$.\footnote{Previously called EFR/EAFR; renamed to reflect inverse polynomial operations.}

Experiments span all combinations of state types (Type-A/B), reward schemes (ILF/IQF), and agent counts ($n \in \{2,3,5,8,10\}$), yielding 20 configurations. Code is available at \url{https://github.com/dentros/Alternation} (GNU GPL). An archived version is available at \url{https://doi.org/10.5281/zenodo.18528891}.

\subsection{Q-Learning and Training}

Each agent maintains an independent tabular Q-table with no inter-agent communication. Hyperparameters: $\gamma = 0.999$, $\alpha = 0.3$, $\epsilon$-greedy with linear decay from $\epsilon_{initial} = 0.9$ to $\epsilon_{min} = 0.004$. Q-learning was selected over deep RL for complete interpretability of coordination dynamics.\footnote{A deep Q-network (DQN) implementation with experience replay was tested under the identical $n=3$, Type-A, ILF configuration (4{,}721 episodes, same schedule): CALT $=0.126$, EALT $=0.197$, FALT $=0.460$, AALT $=0.112$, versus $0.134$, $0.200$, $0.475$, $0.112$ for tabular Q-learning and $0.359$, $0.545$, $0.630$, $0.323$ for the random baseline. DQN performs comparably to or slightly worse than tabular Q-learning, and both remain far below chance, consistent with the non-stationarity induced by concurrent independent learners.} This choice is deliberate. Tabular Q-learning is fully transparent (policies, value estimates, and exploration schedules can be inspected directly; Q-table heatmaps are provided in Supplementary Figures 13 and 14), so the resulting coordination dynamics are interpretable end to end. At the same time, it provides a concrete failure case with diagnostic value: a standard, widely used learning process converges to policies whose coordination failure is invisible to traditional outcome-based metrics yet clearly exposed under ALT evaluation. The contribution lies in the measurement, not in the learner: whether more sophisticated MARL methods can exceed random baselines in this setting remains an open empirical question, and the inadequacy of outcome-based metrics documented here does not depend on its answer.

Episode counts scale with agent complexity:
\begin{equation}
\text{Episodes}(n) = \text{BASE} \times \left(\frac{n}{2}\right)^2 \times \left(1 + \ln\left(\frac{n!}{2!}\right)\right)
\end{equation}
with BASE=1000, yielding: 1,000 (2 agents), 4,721 (3), 31,839 (5), 174,583 (8), and 385,281 (10 agents) episodes per configuration (see Supplementary Section~S5 for full derivation and per-configuration table). This schedule is heuristic, chosen to balance state-space growth against computational and energy constraints; pilot runs with alternative scalings produced qualitatively similar trends.

\subsection{AltRatio and PA-Equivalent Interpretation}

To interpret ALT values, we define $\text{AltRatio} = x/n \in [0,1]$, where $x$ is the equivalent number of perfectly alternating agents. Benchmark sequences are constructed with $x$ agents rotating in fixed order while the remaining $n-x$ agents never reach the terminal, and regression over these integer anchors assigns intermediate (non-integer) equivalents to observed metric values. Multiplying by $n$ yields the \textbf{PA-equivalent}: ``This system coordinates as well as $x$ out of $n$ perfectly alternating agents.'' Benchmarking details and regression equations appear in Supplementary Section S1 (Figures 1 to 6). PA-equivalent is an interpretive mapping distinct from Relative Change and Coordination Score, which compare observed ALT values to random-policy baselines.

\subsection{Random Policy Baselines}\label{sec:random_baselines}

For each configuration, we ran 10,000 episodes with uniformly random actions to establish chance-level performance. Coordination is quantified via:
\begin{equation}
\begin{aligned}
\text{Relative Change} &= \frac{\text{ALT}_{observed} - \text{ALT}_{random}}{\text{ALT}_{random}} \times 100,\\
\text{Coordination Score} &= \frac{\text{ALT}_{observed} - \text{ALT}_{random}}{\text{ALT}_{perfect} - \text{ALT}_{random}} \times 100
\end{aligned}
\end{equation}
where $\text{ALT}_{perfect} = 1.0$. Negative scores indicate performance \textit{worse} than random. Random policies achieve deceptively high traditional metrics (Efficiency $\approx 0.82$, Fairness $\approx 0.97$ for 2 agents) but low ALT values (CALT: 0.486 to 0.111 across agent counts).

The random anchor is not an empirical convention; its level is predicted by probability theory. Under uniform random policies each agent's arrival time is an independent race. In the three-position environment, agent $i$ first reaches the terminal position after $T_i$ rounds, where $P(T_i = t) = (t-1)\,2^{-t}$ for $t \geq 2$ and $P(T_i > t) = (1+t)\,2^{-t}$. An episode ends at $T = \min_i T_i$, and the number of simultaneous winners $K$ has the closed form
\begin{equation}
P(K = k) \;=\; \binom{n}{k} \sum_{t \geq 2} \left[(t-1)\,2^{-t}\right]^{k} \left[(1+t)\,2^{-t}\right]^{n-k},
\label{eq:random_null}
\end{equation}
with winner identities uniform across agents by symmetry, in the spirit of the birthday problem: the probability of an exclusive winner, $P(K{=}1) = 0.815$, $0.681$, $0.483$, $0.283$, $0.193$ for $n = 2, 3, 5, 8, 10$, decays as collisions become the typical outcome. Empirical winner-count frequencies in the 10{,}000-episode random runs match these values to within Monte Carlo error (largest deviation $0.010$, at $n=8$). Because episodes under random play are independent and identically distributed, the expected value of an ALT metric under this null follows by direct evaluation of the episode model of Eq.~\eqref{eq:random_null}. We carry out this evaluation for CALT, the primary metric. The induced null CALT is $0.481$, $0.360$, $0.244$, $0.150$, $0.110$, against measured values of $0.486$, $0.359$, $0.243$, $0.147$, $0.111$. The construction applies unchanged to the other two primary metrics, and we use their measured baselines directly (EALT $0.651 \to 0.183$ and AALT $0.435 \to 0.065$ as $n$ runs from $2$ to $10$), since the CALT agreement above is what establishes that the simulated null reproduces the analytic one. The chance floor used for CALT is therefore theoretically derived rather than post hoc.

This floor also fixes how ALT values should be read as $n$ grows. Chance alone produces real temporal dispersion: uniform randomness scatters wins across agents and episodes, so a moderate ALT value (CALT $\approx 0.49$ at $n=2$) indicates nothing beyond chance, and the floor itself decays with $n$ in the predictable way given above. Raw ALT values are consequently meaningful only relative to the floor at the same $n$, and comparisons across agent counts require the normalized quantities defined above (Relative Change, Coordination Score) or PA-equivalents (Table~\ref{tab:alt_ratio}), since raw gaps shrink mechanically as both learner and floor approach zero. Finally, the floor separates two qualitatively different outcomes: a learner between the floor and PA has achieved partial but genuine coordination, whereas a learner below the floor concentrates access more strongly than chance, destroying the dispersion that randomness provides for free. All Q-learning configurations tested here fall in the second regime.

Falling below the floor is not a foregone conclusion for a deterministic policy, which is what makes the comparison informative rather than definitional. The chance level is not a ceiling on what determinism can achieve, but a midpoint: Perfect Alternation is itself a deterministic policy, and it attains $\mathrm{ALT}=1$, far above the floor at every $n$. A deterministic learner therefore has the full range above the floor available to it, and the round-robin policy that would reach the top of that range is representable in both state encodings used here. Sub-chance performance reflects the particular structure the learner converges to, namely persistent concentration of access, and not the mere fact that its policy is deterministic.

\subsection{Declaration}\label{sec:methods_declaration}

AI-based language models (Claude, ChatGPT) assisted with code organization and language editing during preparation of this study. No experimental design, learning algorithm, or reported result was generated by these tools, and they are not listed as authors. Full disclosure of the specific tools and models used is provided in the Declaration of Generative AI and AI-assisted Technologies section at the end of this manuscript.

\section{Results}\label{sec:results}

\begin{figure}[t]
\centering
\includegraphics[width=\columnwidth]{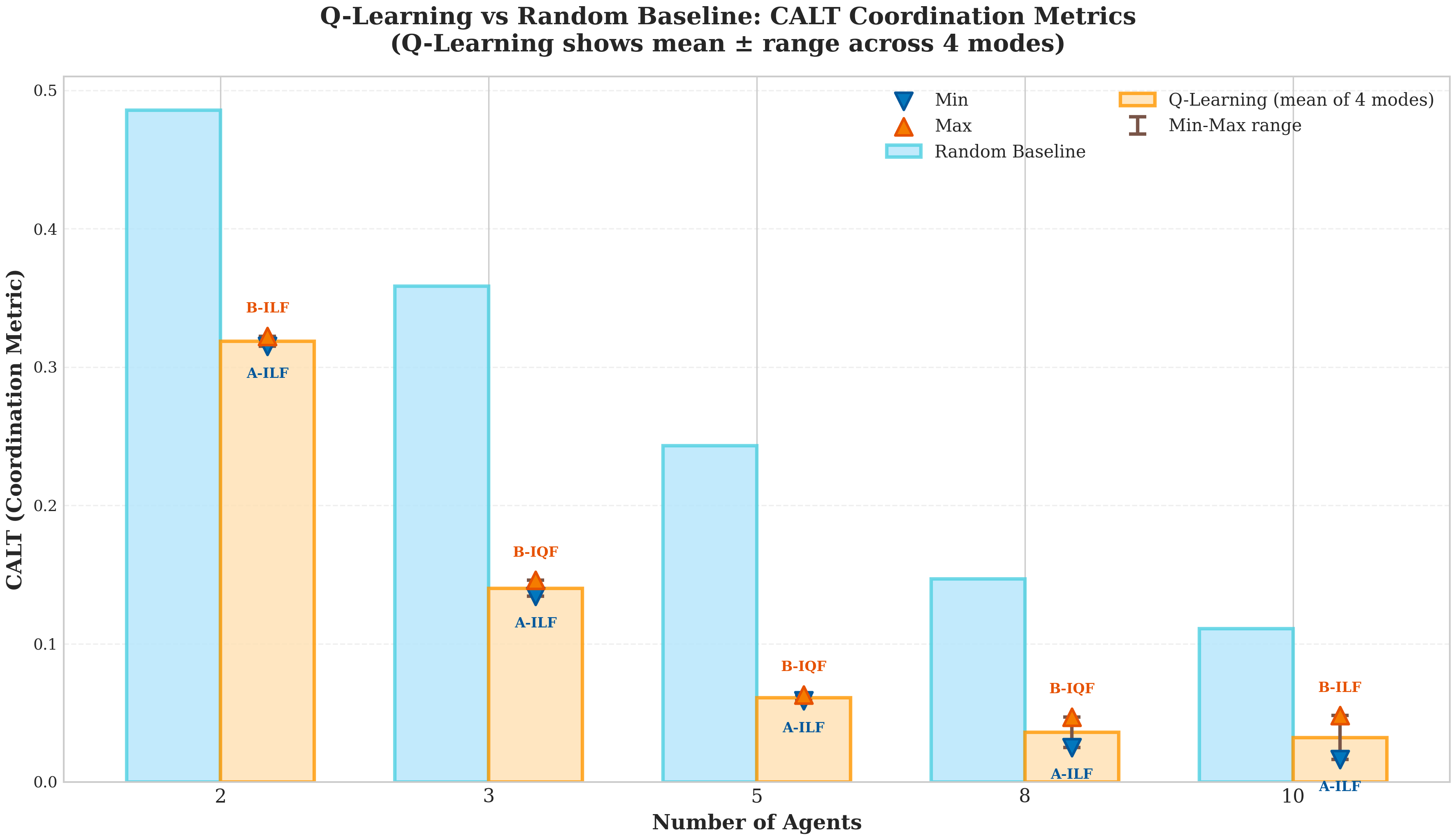}
\caption{CALT for Q-learning (orange bars; mean across the four Type-A/B $\times$ ILF/IQF modes, with min--max range markers) versus the random baseline (light blue bars). Q-learning falls below the chance level at every agent count.}
\label{fig:qlearning_vs_random}
\end{figure}

\begin{figure}[t]
\centering
\includegraphics[width=\columnwidth]{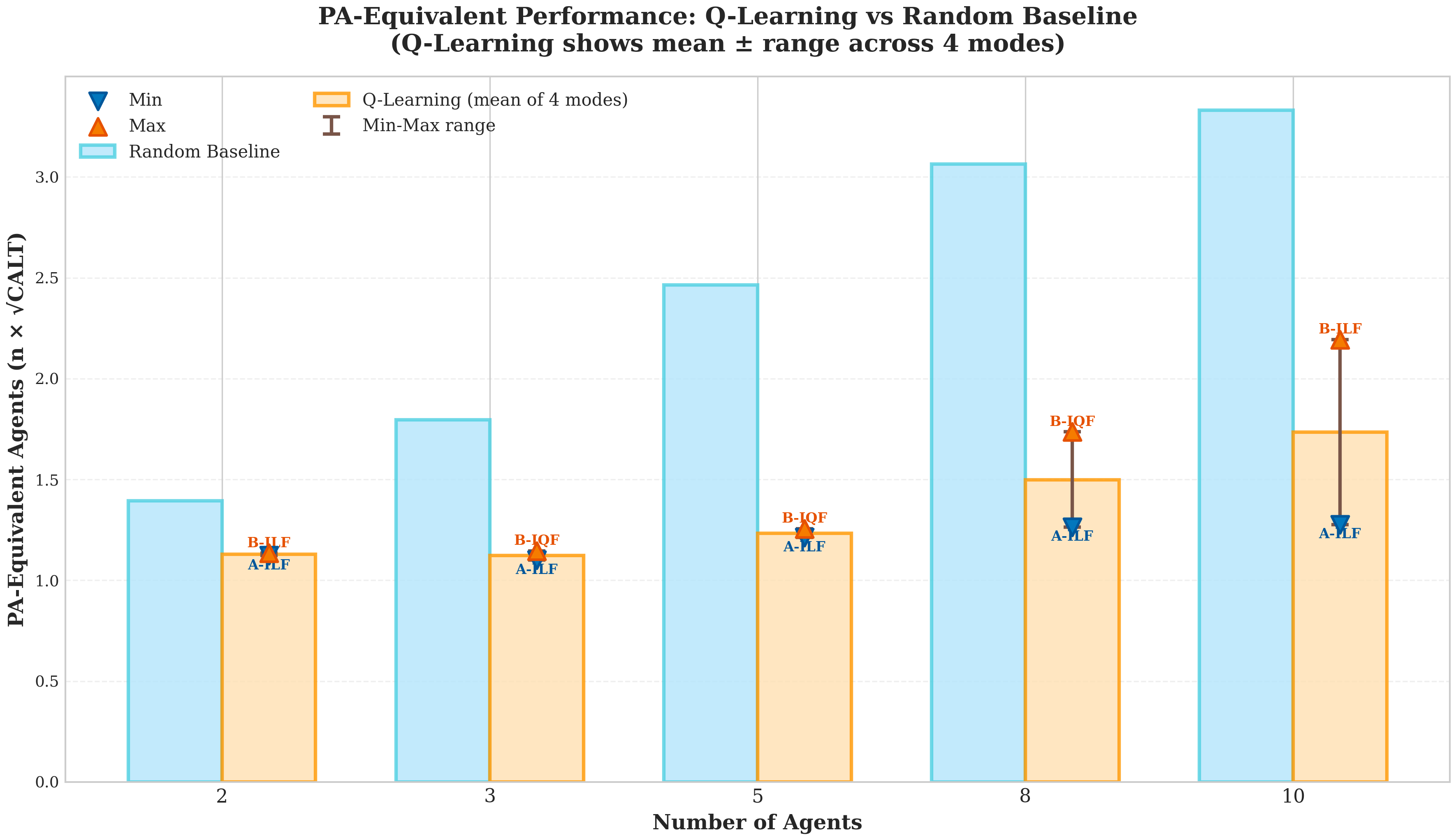}
\caption{PA-equivalent agents ($n \times \sqrt{\mathrm{CALT}}$) for Q-learning (orange bars; mean across the four modes, with min--max range) versus random baselines (light blue). As $n$ grows from 2 to 10, Q-learning rises only from 1.13 to 1.74 equivalent agents (17\% of the system at $n=10$), remaining below the random baseline throughout.}
\label{fig:pa_equivalent}
\end{figure}

\begin{figure*}[t]
\centering
\includegraphics[width=\textwidth]{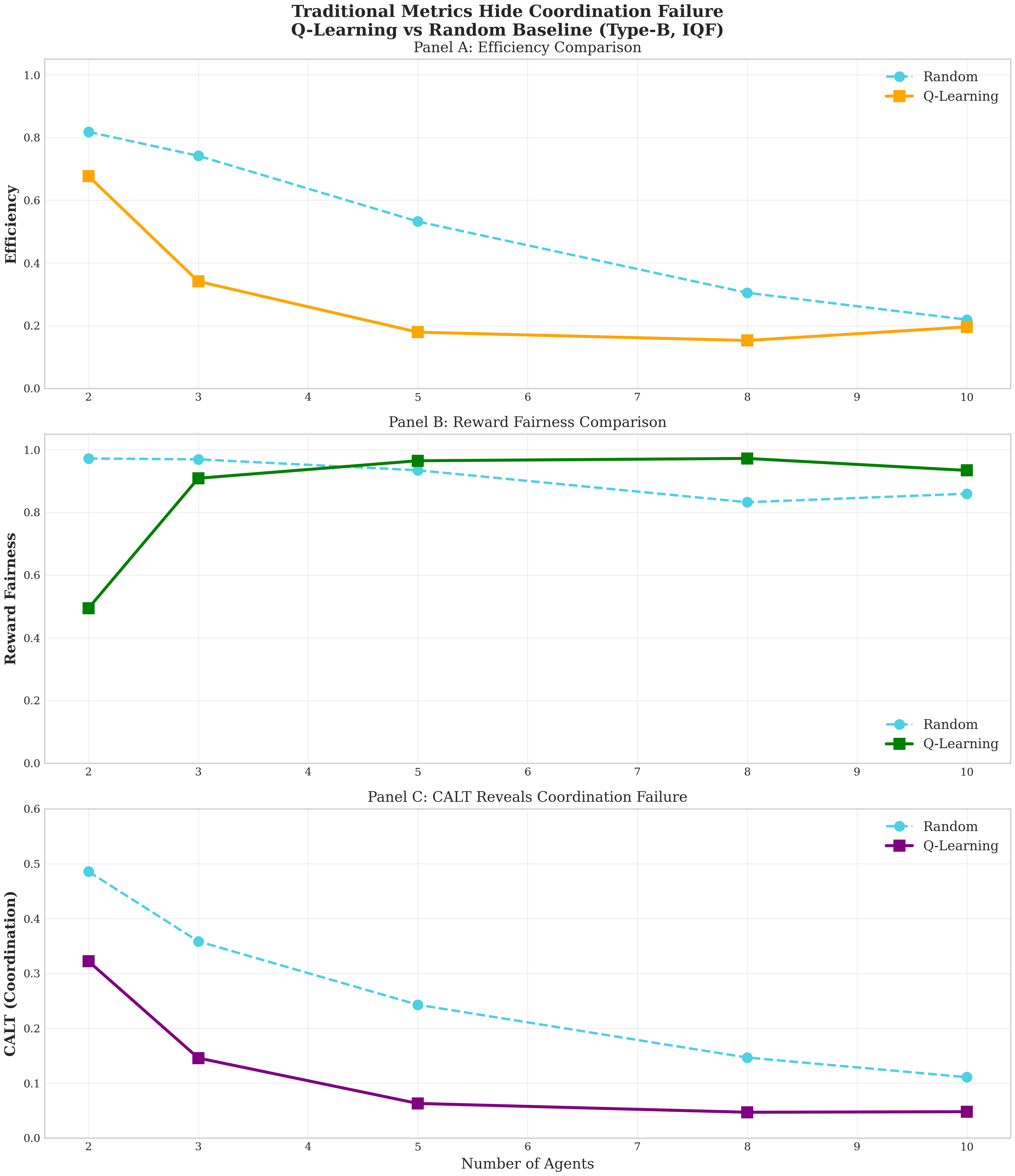}
\caption{Traditional metrics hide coordination failure (Type-B, IQF). Panels top to bottom: Efficiency, Reward Fairness, and CALT, for Q-learning (solid lines) versus random baselines (dashed light blue). Reward Fairness stays above $0.9$ for $n \geq 3$ while Efficiency declines for both processes; only the CALT panel reveals that Q-learning lies consistently below the random baseline.}
\label{fig:traditional_vs_alt}
\end{figure*}

\begin{figure*}[t]
\centering
\includegraphics[width=\textwidth]{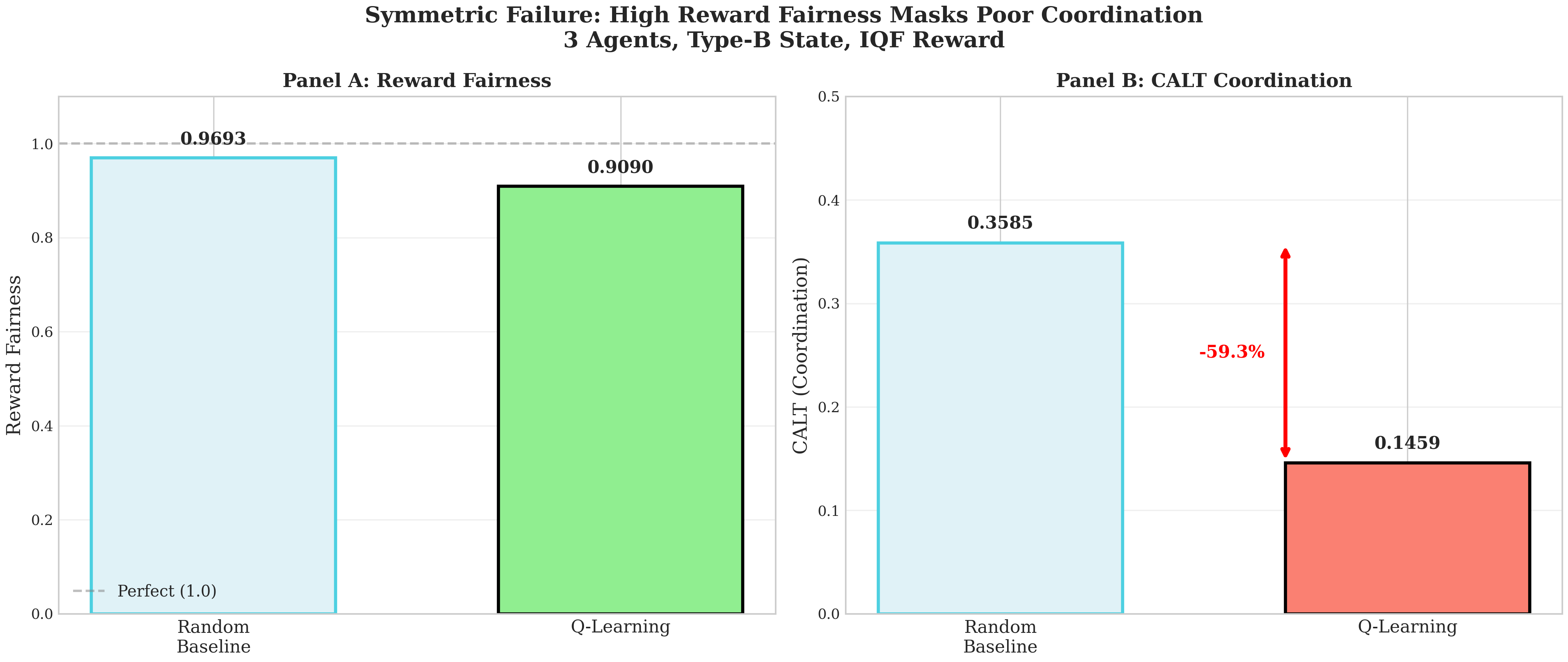}
\caption{Coordination failure at $n=3$ (Type-B, IQF): Reward Fairness remains high for both Q-learning and the random baseline (0.909 vs.\ 0.969), while CALT exposes a severe deficit for Q-learning relative to random (0.146 vs.\ 0.359, a $-59.3\%$ relative drop), showing that a high, seemingly successful Reward Fairness value can co-occur with coordination well below chance.}
\label{fig:symmetric_failure}
\end{figure*}

\begin{figure*}[t]
\centering
\includegraphics[width=\textwidth]{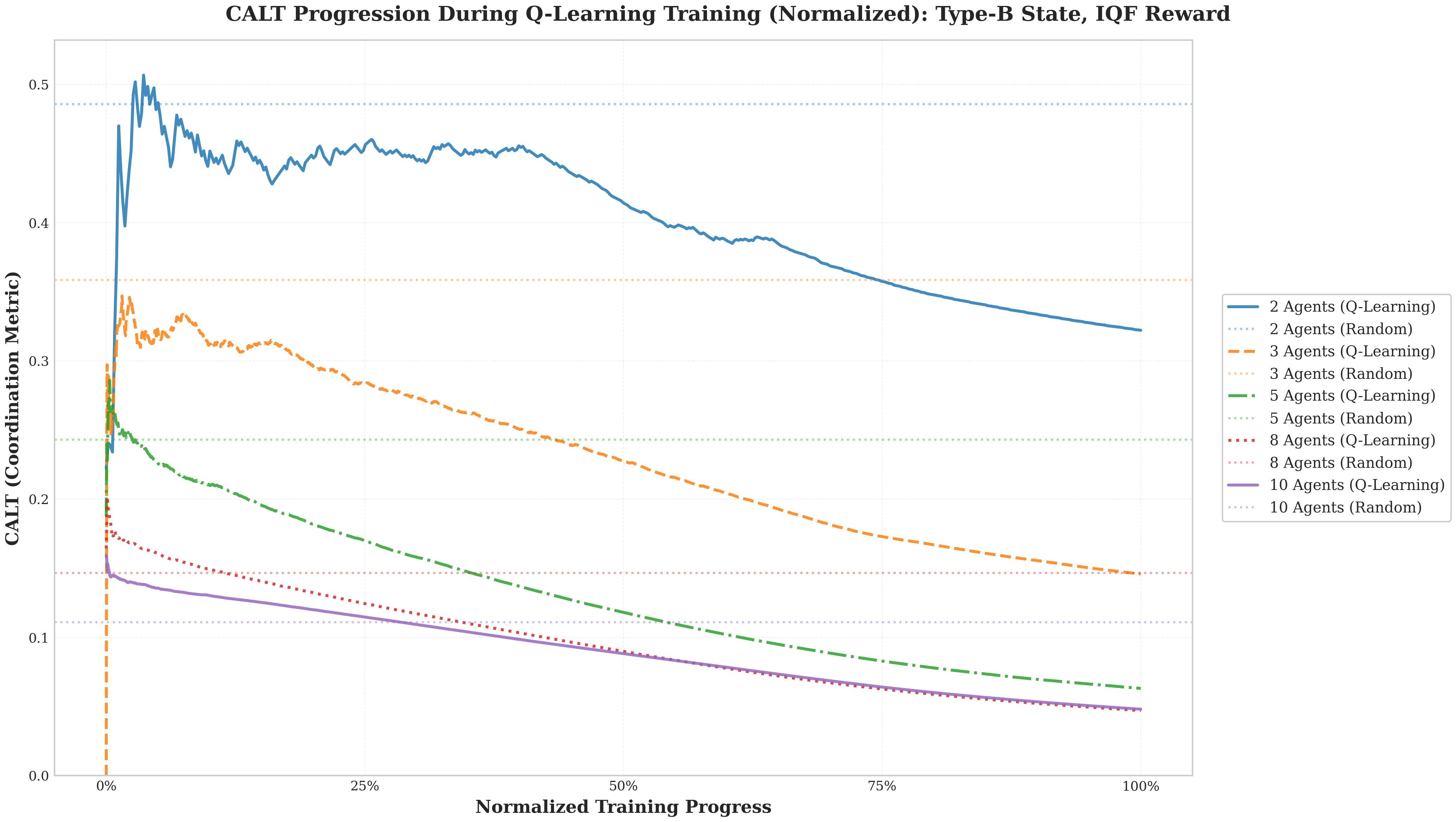}
\caption{CALT progression during Q-learning training (Type-B, IQF) for all five agent counts, with training progress normalized to $[0,1]$ episodes and each agent count's random-policy baseline shown as a horizontal dotted line of matching color. Every curve rises during early exploration and then declines as $\epsilon$ decays toward exploitation, falling below its own random baseline well before training ends, so that learned policies converge to \textit{worse} coordination than the early exploration phase achieved.}
\label{fig:learning_phases}
\end{figure*}

\subsection{Random Baseline Performance}

Table~\ref{tab:random_baselines} presents comprehensive random baseline results across all agent configurations. The key finding is that \textbf{random policies achieve deceptively high traditional metrics}, creating an illusion of coordination:

\begin{table}[t]
\centering
\caption{Random Policy Baseline Performance (10,000 episodes per configuration; averaged across Type-A/B state representations). CALT, FALT, EALT, and Fairness are identical across ILF/IQF reward types: these metrics measure win/loss patterns independent of reward magnitudes. Only Efficiency reflects the reward structure, and from $n=3$ onward it declines more sharply under IQF (tie reward $= r_{\mathrm{high}}/n^2$) as multi-agent ties become more frequent with $n$. Despite producing no temporal coordination, random policies achieve Fairness $\geq 0.816$ and Efficiency $\geq 0.219$ across all configurations.}
\label{tab:random_baselines}
\small
\resizebox{\columnwidth}{!}{%
\begin{tabular}{cccccc}
\hline
\textbf{Agents} & \textbf{CALT} & \textbf{FALT} & \textbf{EALT} & \textbf{Efficiency (ILF / IQF)} & \textbf{Fairness} \\
\hline
2  & 0.486 & 0.717 & 0.651 & 0.818 / 0.818 & 0.972 \\
3  & 0.359 & 0.630 & 0.545 & 0.866 / 0.742 & 0.967 \\
5  & 0.243 & 0.526 & 0.410 & 0.727 / 0.532 & 0.927 \\
8  & 0.147 & 0.416 & 0.251 & 0.526 / 0.305 & 0.816 \\
10 & 0.111 & 0.363 & 0.183 & 0.443 / 0.219 & 0.846 \\
\hline
\end{tabular}%
}
\end{table}

For 2-agent systems, random policies achieve Efficiency $= 0.818$ and Fairness $= 0.972$, values that would typically be read as successful coordination, and even with 5 agents they maintain Efficiency $\geq 0.532$ and Fairness $\geq 0.927$, despite the complete absence of any learned strategy. ALT metrics expose the difference: random CALT decreases monotonically from 0.486 (2 agents) to 0.111 (10 agents), tracking the analytically predicted chance floor of Section~\ref{sec:random_baselines}.

Figure~\ref{fig:qlearning_vs_random} visualizes this comparison, showing that Q-learning agents (orange bars) consistently achieve \textit{lower} CALT values than random baselines (light blue bars) across all agent configurations, a finding that challenges assumptions about learned coordination in this class of multi-agent systems. We emphasize that this result does not generalize to all MARL algorithms, but highlights a failure mode of independent tabular learners under temporal coordination demands. We report CALT explicitly as the primary alternation metric; other ALT variants are summarized in Table~\ref{tab:coordination_analysis} and, across the full range of agents and modes tested, in Supplementary Figures 17 and 18.

\subsection{Q-Learning Coordination Failure}

Our Q-learning experiments reveal that, despite achieving high traditional fairness and efficiency metrics, Q-learning agents demonstrate \textbf{systematically worse coordination than random policies} in this setting. Table~\ref{tab:coordination_analysis} reports both \textbf{Relative Change} and \textbf{Coordination Score} for FALT, EALT, CALT, and AALT:

\begin{table}[t]
\centering
\caption{Q-Learning vs Random Baseline: Coordination Score Analysis (BASE=1000; Type-B state representation). All coordination scores are \textbf{negative}, indicating Q-learning performs worse than random baselines on every primary ALT metric across all agent counts. Coordination scores narrow as $n$ increases because both Q-learning and random ALT values converge toward zero (see Supplementary Section~S6); the magnitude of the gap is best interpreted via Table~\ref{tab:alt_ratio}. \textbf{Relative Change:} $(\text{QL}-\text{Random})/\text{Random}\times100$. \textbf{Coord.\ Score:} $(\text{QL}-\text{Random})/(\text{Perfect}-\text{Random})\times100$.}
\label{tab:coordination_analysis}
\small
\resizebox{\columnwidth}{!}{%
\begin{tabular}{clcccc}
\hline
\textbf{Agents} & \textbf{Metric} & \textbf{Q-Learning} & \textbf{Random} & \textbf{Rel.\ Change (\%)} & \textbf{Coord.\ Score (\%)} \\
\hline
\multirow{4}{*}{2}
 & CALT  & 0.322 & 0.486 & \textcolor{red}{-33.7\%} & \textcolor{red}{-31.8\%} \\
 & EALT  & 0.513 & 0.651 & \textcolor{red}{-21.2\%} & \textcolor{red}{-39.6\%} \\
 & AALT  & 0.284 & 0.435 & \textcolor{red}{-34.8\%} & \textcolor{red}{-26.8\%} \\
 & FALT  & 0.642 & 0.717 & \textcolor{red}{-10.5\%} & \textcolor{red}{-26.8\%} \\
\hline
\multirow{4}{*}{3}
 & CALT  & 0.142 & 0.359 & \textcolor{red}{-60.3\%} & \textcolor{red}{-33.7\%} \\
 & EALT  & 0.203 & 0.545 & \textcolor{red}{-62.7\%} & \textcolor{red}{-75.0\%} \\
 & AALT  & 0.115 & 0.323 & \textcolor{red}{-64.3\%} & \textcolor{red}{-30.6\%} \\
 & FALT  & 0.480 & 0.630 & \textcolor{red}{-23.8\%} & \textcolor{red}{-40.6\%} \\
\hline
\multirow{4}{*}{5}
 & CALT  & 0.062 & 0.243 & \textcolor{red}{-74.3\%} & \textcolor{red}{-23.8\%} \\
 & EALT  & 0.099 & 0.410 & \textcolor{red}{-75.9\%} & \textcolor{red}{-52.7\%} \\
 & AALT  & 0.043 & 0.203 & \textcolor{red}{-78.8\%} & \textcolor{red}{-20.1\%} \\
 & FALT  & 0.313 & 0.526 & \textcolor{red}{-40.6\%} & \textcolor{red}{-45.0\%} \\
\hline
\multirow{4}{*}{8}
 & CALT  & 0.047 & 0.147 & \textcolor{red}{-68.2\%} & \textcolor{red}{-11.7\%} \\
 & EALT  & 0.109 & 0.251 & \textcolor{red}{-56.6\%} & \textcolor{red}{-19.0\%} \\
 & AALT  & 0.035 & 0.100 & \textcolor{red}{-65.0\%} & \textcolor{red}{-7.2\%} \\
 & FALT  & 0.240 & 0.416 & \textcolor{red}{-42.2\%} & \textcolor{red}{-30.0\%} \\
\hline
\multirow{4}{*}{10}
 & CALT  & 0.048 & 0.111 & \textcolor{red}{-56.6\%} & \textcolor{red}{-7.1\%} \\
 & EALT  & 0.166 & 0.183 & \textcolor{red}{-9.5\%} & \textcolor{red}{-2.1\%} \\
 & AALT  & 0.042 & 0.065 & \textcolor{red}{-35.1\%} & \textcolor{red}{-2.5\%} \\
 & FALT  & 0.232 & 0.363 & \textcolor{red}{-36.2\%} & \textcolor{red}{-20.6\%} \\
\hline
\end{tabular}%
}
\end{table}

\textbf{Key Finding:} For FALT/EALT/CALT/AALT, all \textbf{Coordination Scores} are negative, indicating performance below random baselines after normalization to perfect alternation. Relative Change is negative for all ALT variants except 10-agent Type-B qEALT ($+5.4\%$ under ILF, $+7.4\%$ under IQF), where more frequent multi-agent ties yield non-zero rewards and behavior drifts toward the random baseline; the quadratic form amplifies small relative differences and tracks the slightly higher efficiency in that regime (see Supplementary Section~S2).

Across the primary metrics reported in Table~\ref{tab:coordination_analysis}, Relative Change spans from $-9.5\%$ (EALT, 10 agents) to $-78.8\%$ (AALT, 5 agents), and Coordination Scores range from $-2.1\%$ to $-75.0\%$. Including secondary quadratic variants (detailed in Supplementary Section~S1), the worst observed Relative Change for Type-B reaches $-81.2\%$ for qEALT at 5 agents. For Type-A (position-only) states, the deficit is more severe and monotonically increasing: EALT Relative Change reaches $-92.2\%$ at $n=10$ (vs.\ $-9.5\%$ for Type-B at the same $n$), confirming that episodic memory substantially mitigates coordination failure at large $n$ while position-only representations show relentless degradation (see Supplementary Table~S1 for full comparison). The most severe cases among primary metrics are:
\begin{itemize}
\item \textbf{5 agents, AALT:} Relative Change $= -78.8\%$; Coordination Score $= -20.1\%$
\item \textbf{3 agents, EALT:} Coordination Score $= -75.0\%$ (worst normalized gap)
\item \textbf{10 agents, CALT:} Relative Change $= -56.6\%$ (severe degradation at scale)
\end{itemize}

The non-monotonic profile of Relative Change across $n$ (worst near $n=5$) reflects two opposing forces: coordination difficulty grows with $n$, while the chance floor itself collapses (Section~\ref{sec:random_baselines}), mechanically compressing relative gaps at large $n$.

Meanwhile, traditional metrics mask this failure at precisely the experiments where coordination is worst. The maximum contrast among primary metrics occurs at 5 agents (AALT Relative Change $= -78.8\%$) and 3 agents (EALT Coordination Score $= -75.0\%$). In these same configurations traditional metrics still look benign: at $n=5$ Reward Fairness is $0.962$ and TT Fairness $0.984$, while at $n=3$ the win-count Fairness reaches $0.992$ (Supplementary Table~S1), values that would conventionally be interpreted as near-perfect coordination. Our ALT metrics reveal the opposite: Q-learning agents have learned policies that are \textit{actively worse} than random action selection for achieving turn-taking coordination in this setting.

Figure~\ref{fig:traditional_vs_alt} illustrates this dichotomy, with high traditional metrics alongside low ALT scores across all tested configurations.

\subsection{Perfect Alternation Equivalent Analysis}

To provide interpretable quantification of coordination quality, we compute PA-equivalent performance using the AltRatio framework. Table~\ref{tab:alt_ratio} presents this analysis for CALT:

\begin{table}[t]
\centering
\caption{Perfect Alternation Equivalent Analysis (CALT, BASE=1000; Type-B state representation). $\text{ALT Ratio} = \sqrt{\text{CALT}}$ (the fitted intercept of the benchmarking regression, $\sim\!2\times10^{-10}$, is a numerical-precision artifact and is treated as zero); $\text{PA Equiv.} = n \times \text{ALT Ratio}$. IQF (tie reward $= r_{\mathrm{high}}/n^2$) consistently yields slightly higher CALT than ILF (tie reward $= r_{\mathrm{high}}/n$) at intermediate $n$: the lower tie incentive under IQF exerts marginally stronger pressure toward exclusive wins, improving coordination structure. However, the effect is small ($\Delta \leq 0.004$) and both reward structures produce a sharp degradation from $\sim$57\% at $n=2$ to $\sim$21\% at $n\geq8$, confirming that coordination failure is reward-structure-agnostic.}
\label{tab:alt_ratio}
\resizebox{\columnwidth}{!}{%
\begin{tabular}{ccccccc}
\hline
\textbf{Agents} & \textbf{Reward} & \textbf{CALT (QL)} & \textbf{ALT Ratio} & \textbf{PA Equiv.} & \textbf{Performance} \\
 & \textbf{Type} & & & \textbf{Agents} & \textbf{(\% of Perfect)} \\
\hline
\multirow{2}{*}{2}  & ILF & 0.3222 & 0.568 & 1.14 & 56.8\% \\
                    & IQF & 0.3222 & 0.568 & 1.14 & 56.8\% \\
\hline
\multirow{2}{*}{3}  & ILF & 0.1423 & 0.377 & 1.13 & 37.7\% \\
                    & IQF & 0.1459 & 0.382 & 1.15 & 38.2\% \\
\hline
\multirow{2}{*}{5}  & ILF & 0.0624 & 0.250 & 1.25 & 25.0\% \\
                    & IQF & 0.0631 & 0.251 & 1.26 & 25.1\% \\
\hline
\multirow{2}{*}{8}  & ILF & 0.0467 & 0.216 & 1.73 & 21.6\% \\
                    & IQF & 0.0471 & 0.217 & 1.74 & 21.7\% \\
\hline
\multirow{2}{*}{10} & ILF & 0.0482 & 0.219 & 2.19 & 21.9\% \\
                    & IQF & 0.0481 & 0.219 & 2.19 & 21.9\% \\
\hline
\end{tabular}%
}
\end{table}

Table~\ref{tab:alt_ratio} reports results for the Type-B state representation (ILF and IQF shown separately); Figure~\ref{fig:pa_equivalent} visualizes the mean $\pm$ range across all four modes (Type-A/B $\times$ ILF/IQF). For $n \leq 5$, both state types yield nearly identical CALT values; for $n \geq 8$, Type-B yields higher absolute values due to episodic memory (see Supplementary Table~S1), contributing to the wider shaded intervals at large $n$. The corresponding PA-equivalent analysis for the remaining five ALT variants is provided in Supplementary Figures 19 to 24.

Performance drops sharply from 56.8\% of perfect coordination at 2 agents to 25.0\% at 5, and remains near $\sim$22\% for 8 to 10 agents (Table~\ref{tab:alt_ratio}; Figure~\ref{fig:pa_equivalent}). Notably, 10 Q-learning agents achieve coordination equivalent to only 2.19 perfectly alternating agents, barely one-fifth of the system exhibiting coordinated behavior.

\subsection{Symmetric Coordination Failure}

A particularly illuminating case study emerges from comparing 3-agent configurations across state representations and reward structures. Under Type-B, IQF (Figure~\ref{fig:symmetric_failure}), Reward Fairness for Q-learning remains high relative to random (0.909 vs.\ 0.969) even as CALT exposes a severe deficit (0.146 vs.\ 0.359, a $-59.3\%$ relative drop). The same pattern holds under Type-A, ILF, where Reward Fairness is likewise high (0.921) alongside a comparably low CALT ($\approx 0.134$). The disagreement can run the other way, too: under the ILF reward alone, Reward Fairness diverges sharply between state types (0.921 for Type-A versus 0.640 for Type-B, driven by one agent participating in substantially fewer partial-tie episodes despite winning exclusively at essentially the same rate as its peers; Supplementary Section~S2), while win-count Fairness ranks the two state types in the opposite order (0.898 versus 0.992); CALT is unmoved by either disagreement, again $\approx 0.13$--$0.14$ for both. Whether traditional metrics agree or actively contradict each other across configurations, CALT identifies the same coordination failure, indicating that the deficit is not an artifact of any one experimental design choice within HJG, though whether it extends beyond tabular Q-learning remains open (Section~\ref{sec:experiments}). For $n \geq 8$, Type-B state representation yields higher absolute CALT values owing to episodic memory (e.g., CALT$_{\mathrm{B}}=0.047$ vs.\ CALT$_{\mathrm{A}}=0.025$ at $n=8$; see Supplementary Table~S1), though Q-learning remains well below the random baseline in both cases.

\FloatBarrier
\section{Discussion}\label{sec:discussion}

\subsection{Why Does Q-Learning Fail at Coordination?}

Our finding that independent Q-learning agents perform below random baselines demands explanation. We identify four interrelated factors:

\textbf{1. The Credit Assignment Problem.} Tabular Q-learning cannot recognize that ``losing now'' (allowing another agent to win) yields ``winning later'' (reciprocal cooperation): the reward for cooperation materializes $n$ episodes in the future, beyond the effective credit-assignment horizon of independent tabular learners even with $\gamma = 0.999$.

\textbf{2. Non-Stationary Opponent Dynamics.} From each agent's perspective, other agents' evolving policies violate the stationarity assumptions underlying Q-learning convergence. Once all agents simultaneously cease exploration (synchronized epsilon decay at 75\% of episodes), they may lock into a suboptimal, locally stable regime where no single agent can unilaterally improve by alternating~\citep{masiliunas2019}; the same synchronized decay likely produces the mid-range bumps in EALT/qEALT, as partial exploration increases exclusive wins before late exploitation drives simultaneous arrivals.

\textbf{3. Lack of Explicit Coordination Signals.} Unlike humans who achieve PA through implicit signaling (e.g., ``I won last time, your turn now''), independent learners have no communication channel. State representations lack sufficient context to infer ``whose turn it is,'' particularly for $n > 3$ agents where alternation cycles exceed the single-episode memory of Type-B states.

\textbf{4. Tragedy of the Learning Commons.} Each agent learns to exploit opportunities for $r_{high}$ greedily, maximizing individual expected return. Without global coordination mechanisms, this individual rationality leads to collective irrationality, as agents interfere with each other's learning trajectories, producing worse outcomes than random policies that at least distribute wins stochastically.

These factors compound as agent count increases, producing the degradation profile of Table~\ref{tab:alt_ratio}. The coordination problem is not merely difficult; it is poorly aligned with the assumptions of independent tabular reinforcement learning.

\subsection{Coordination Breakdown as Strategic Regime Shift}

The degradation with agent count reflects a shift in the equilibrium structure of the game: as $n$ grows, the strategic complexity of coordinated alternation increases sharply, and independent learners fail to discover equilibria supporting fair turn-taking~\citep{pikovsky2009,maruta2012,masiliunas2019}. That 10 Q-learning agents coordinate as effectively as only 2.19 perfectly alternating agents (Type-B; 1.28 for Type-A) identifies a regime of near-complete coordination breakdown, information that remains hidden under traditional metrics whose values can appear deceptively high (Reward Fairness up to 0.993, Efficiency up to 0.677). Performance stays close to random between 8 and 10 agents, with a slight uptick in Type-B EALT at $n=10$ (episodic memory provides marginal advantage at scale), suggesting a near-random regime that merits finer-grained agent-count sweeps. Across this range, CALT separates perfect alternation, learned partial alternation, and random access cleanly by construction (Section~\ref{sec:alt}), but the empirical regimes observed here collapse toward the low end of that scale: none of the tested configurations reaches the intermediate, genuinely coordinated band between the random floor and PA, so the discriminative power the metric offers in principle is not exercised by the Q-learning policies actually produced.

\subsection{Implications for Random Baseline Methodology}

The random baseline supplies interpretive weight that AltRatio alone cannot: knowing that CALT $= 0.047$ corresponds to 21.6\% of perfect alternation does not reveal whether this exceeds or falls below chance-level behavior. Because the anchor is analytically derived (Section~\ref{sec:random_baselines}), falling below it is the strongest form of failure: learned deterministic policies concentrate access more strongly than chance would, consistent with the learning-phase analysis of Figure~\ref{fig:learning_phases}, where CALT declines precisely as exploration decays; per-configuration learning curves and epsilon trajectories are provided in Supplementary Figures 7 to 12, and rolling-window terminal-win patterns underlying this analysis in Supplementary Figures 15 and 16. Additional discussion on interpretability and reconciliation with prior work is provided in Supplementary Section S6.

\FloatBarrier
\section{Conclusions}\label{sec:conclusion}

This study introduces Perfect Alternation (PA) as a reference coordination regime and limiting benchmark for temporal turn-taking, alongside six novel ALT metrics and a regression-based benchmarking framework enabling interpretable quantification of coordination quality. Through comprehensive Q-learning experiments with rigorous random policy baselines (an approach rarely formalized as an explicit null process in prior BoE work), we reveal fundamental limitations in both traditional evaluation metrics and learned coordination strategies~\citep{hawkins2016,puig2018,gasparrini2018,freire2020,freire2023}.

\subsection{Key Findings}

Our experimental results challenge conventional assumptions about multi-agent coordination in three critical ways:

\textbf{1. Traditional Metrics Do Not Detect Coordination Failure.} Across all configurations, Q-learning agents exhibit Reward Fairness up to 0.993 and TT Fairness up to 0.998, values that appear deceptively high, while ALT Relative Change reaches as low as $-92\%$ (EALT, Type-A at $n=10$), with a single small positive at $+5.4\%$ (10-agent qEALT). Outcome-based fairness measures cannot distinguish monopolistic, random, or alternating access patterns in temporal multi-agent scenarios.

\textbf{2. Q-Learning Produces Systematic Un\-der-Co\-or\-di\-na\-tion in the Honey-Jar Game.} With that one negligible exception, \textit{all} coordination scores are negative, a result not previously established in the BoE literature. Independent Q-learning agents do not merely fail to coordinate; they learn policies that suppress alternation. CALT Relative Change reaches $-33.7\%$ at $n=2$, $-74.3\%$ at $n=5$, and $-56.6\%$ at $n=10$, with Coordination Scores of $-31.8\%$, $-23.8\%$, and $-7.1\%$ respectively, a pattern confirmed across the full ALT strictness spectrum.

\textbf{3. Coordination Difficulty Scales Sharply with Agent Count.} PA-equivalent performance drops from 56.8\% (2 agents) to 25.0\% (5 agents); with episodic memory (Type-B) it plateaus near $\sim$22\% for 8 to 10 agents, while position-only states (Type-A) decline monotonically to $\sim$12.8\% at $n=10$. Both representations remain well below random baselines throughout, so episodic memory mitigates but does not resolve the deficit (Supplementary Table~S1).

\subsection{Implications, Limitations, and Future Directions}

Our results highlight constraints of independent tabular learning for temporal coordination and point to promising directions, including richer state representations, explicit communication mechanisms, and centralized training with decentralized execution. Extended implications, limitations, future directions, and application contexts are provided in Supplementary Section S6. Results are qualitatively consistent across all four configurations (Type-A/B $\times$ ILF/IQF): all coordination scores remain negative regardless of state type or reward scheme. For $n \leq 5$, absolute CALT values are nearly identical across state types; for $n \geq 8$, Type-B yields higher absolute values due to episodic memory, though coordination failure persists in both (Supplementary Table~S1). Coordination failure is therefore not an artifact of any specific state representation or reward structure. Two further limitations bound the scope of the learning results: all experiments use a single hyperparameter configuration (one exploration schedule, no eligibility traces), so the Q-learning outcomes are a documented failure case rather than a claim about tabular learning in general; and all measurements come from one game family, although the metric definitions are game-agnostic and their validity rests on constructed sequences and probability theory rather than on this testbed (Section~\ref{sec:alt}). On the computational side, the ALT family evaluates all $b = \nu - (n-1)$ windows in $O(\nu n)$ time, which is practical at the scales studied here but grows heavy for large populations; a companion line of work introduces Rotational Periodicity, a lightweight $O(\nu + n)$ measure family that tracks the ALT metrics closely and can serve as their scalable proxy~\citep{papadopoulos2025}. A full axiomatic characterization of the ALT family, in the spirit of axiomatizations of queueing fairness measures~\citep{avi-itzhak2008}, establishing which normative properties uniquely determine CALT, or revealing that none do, is left for future theoretical work.

Although developed on HJG, the framework applies unchanged to any setting in which $n$ agents repeatedly contend for a single indivisible resource slot: wireless spectrum access ($n$ devices, one channel per slot)~\citep{shenoy2007}, traffic intersection control ($n$ approaches, one green phase per cycle)~\citep{wei2021}, robotic task allocation ($n$ robots, one task per timestep)~\citep{mezey2025}, and compute-cluster job scheduling ($n$ jobs, one execution slot)~\citep{mao2016}. In each case the sequence of exclusive winners is directly observable, so ALT metrics, PA-equivalent benchmarking, and random-baseline comparison can be computed without modification. For settings with graded rather than binary success, defining the per-episode winner as the agent obtaining the maximal reward extends the framework to measuring temporal equity of reward dominance.

More broadly, our findings show that distributional fairness criteria, widely used in fair division and cooperative game theory, can be misleading in repeated interaction settings where the \emph{temporal sequence} of allocations matters as much as their aggregate distribution. Temporal fair division \citep{cookson2025,elkind2025,choi2026}, picking sequence analysis \citep{bouveret2014,kalinowski2013}, perpetual voting \citep{lackner2020}, and temporally aware optimization \citep{torres2024} all show that cumulative criteria alone are insufficient; our results complement this line of work from a diagnostic perspective, asking whether self-interested agents can achieve temporal fairness spontaneously through reinforcement learning and quantifying the gap when they do not. The difficulty of the mechanism design problem itself motivates the diagnostic approach: \citet{elkind2025} prove that determining whether a TEF1 allocation exists is NP-hard in general, and \citet{choi2026} show that strict temporal fairness remains impossible in most settings even with a scheduling buffer, so empirical emergence cannot be assumed and must instead be measured. \citet{choi2026} further identify strategy-proofness under competitive settings as an open question; our results provide an empirical answer for the reinforcement learning analogue, with Relative Change spanning roughly $-34\%$ to $-79\%$ (CALT and AALT, Type-B) and reaching $-92\%$ for EALT in position-only states. For comparison, \citet{bouveret2014} establish that even deliberate worst-case strategic manipulation of an alternating picking sequence degrades social welfare by at most $\sim$33\%; that spontaneous coordination failure here exceeds this bound illustrates the broader point: without temporally sensitive metrics, failures of this magnitude would remain invisible, misclassified as successful coordination.


\section*{Declaration of Generative AI and AI-assisted Technologies in the Writing Process}
During preparation of this manuscript, the authors used AI tools (Claude 3.7/4.5,
Claude Sonnet 5; ChatGPT 5.2) for language editing, formatting, and code
organization, and Manus AI for a small part of the literature review. The authors
reviewed and edited all outputs and take full responsibility for the content. The
use of these tools did not influence the scientific findings or conclusions.

\section*{CRediT authorship contribution statement}

\textbf{Nikolaos Al. Papadopoulos:} Conceptualization, Methodology, Software, Formal analysis, Investigation, Data curation, Writing -- original draft, Visualization. \textbf{Ismael Tito Freire:} Conceptualization (AltRatio benchmarking methodology), Supervision, Writing -- review \& editing. \textbf{Marti Sanchez-Fibla:} Conceptualization (BoE-based framework from prior work this study builds on), Validation. \textbf{Konstantinos E. Psannis:} Supervision, Resources.

\section*{Acknowledgments}

All code and experimental infrastructure used in this study were designed and implemented by N. Al. Papadopoulos, who retains ownership of the codebase; it is released under an open GNU GPL license, as noted in the Data Availability statement.

\section*{Declaration of competing interest}

The authors declare that they have no known competing financial interests or personal relationships that could have appeared to influence the work reported in this paper.

\section*{Data availability}

The data that support the findings of this study are openly available in Zenodo at \url{https://doi.org/10.5281/zenodo.18528891}.

\section*{Funding}

This research did not receive any specific grant from funding agencies in the public, commercial, or not-for-profit sectors.

\label{end:maintext}

\end{document}